\documentclass[conference]{IEEEtran}
\IEEEoverridecommandlockouts
\usepackage{cite}
\usepackage{amsmath,amssymb,amsfonts}
\usepackage{algorithmic}
\usepackage{graphicx}
\usepackage{textcomp}
\usepackage{xcolor}

\usepackage{graphicx}

\usepackage{balance}
\usepackage{array}
\usepackage{tabulary}
\newcolumntype{K}[1]{>{\centering\arraybackslash}p{#1}}

\usepackage{diagbox}

\usepackage{relsize}
\usepackage{verbatim}

\usepackage{amsthm}
\usepackage{hyperref}
\usepackage{subfigure}
\usepackage{enumitem}
\usepackage{booktabs} 
\usepackage{multirow}
\usepackage{tabularx}

\usepackage[para]{footmisc}

\usepackage{tikz}
\usetikzlibrary{trees}
\usetikzlibrary{positioning}
\usetikzlibrary{matrix}
\definecolor{myyellow}{RGB}{250, 246, 145}
\definecolor{myorange}{RGB}{245, 203, 130}
\definecolor{mygreen}{RGB}{211, 247, 188}
\definecolor{myblue}{RGB}{86, 180, 233}
\def\BibTeX{{\rm B\kern-.05em{\sc i\kern-.025em b}\kern-.08em
    T\kern-.1667em\lower.7ex\hbox{E}\kern-.125emX}}

\newtheorem*{definition}{Definition}
\newcounter{AsteriosNOC}
\stepcounter{AsteriosNOC}

\newcounter{ChristosNOC}
\stepcounter{ChristosNOC}

\newcounter{ChristophNOC}
\stepcounter{ChristophNOC}

\newcounter{MariosNOC}
\stepcounter{MariosNOC}

\newcounter{AngelaNOC}
\stepcounter{AngelaNOC}

\newcommand{\new}{\textcolor{black}}

\newcommand{\rev}{\textcolor{black}}

\newcommand{\rgt}[0]{recall at ground truth}

\newcommand{\suite}[0]{Valentine}
\newcommand{\para}[1]{\vspace{1mm}\noindent\textbf{#1.}}
\begin{document}

\title{Valentine: Evaluating Matching Techniques for \\ Dataset Discovery}
\author{Christos Koutras$^1$ \hfill George Siachamis$^{1,2}$ \hfill Andra Ionescu$^1$ \hfill Kyriakos Psarakis$^1$ \hfill Jerry Brons$^2$ \vspace{.7mm}\\ \hfill Marios Fragkoulis$^1$ \hfill Christoph Lofi$^1$ \hfill Angela Bonifati$^3$ \hfill Asterios Katsifodimos$^1$ \vspace{2.5mm} \\
$^1$\textit{Delft Univeristy of Technology} ~~ $^2$\textit{ING Bank Netherlands} ~~$^3$\textit{Lyon 1 University}
}

\maketitle

\begin{abstract}
Data scientists today search large data lakes to discover and integrate datasets. In order to bring together disparate data sources, dataset discovery methods rely on some form of schema matching: the process of establishing correspondences between datasets. Traditionally, schema matching has been used to find matching pairs of columns between a source and a target schema. 
However, the use of schema matching in dataset discovery methods differs from its original use. Nowadays schema matching serves as a building block for indicating and ranking inter-dataset relationships.
Surprisingly, although a discovery method's success relies highly on the quality of the underlying matching algorithms, the latest discovery methods employ existing schema matching algorithms in an ad-hoc fashion due to the lack of openly-available datasets with ground truth, reference method implementations, and evaluation metrics.


 In this paper, we aim to rectify the problem of evaluating the effectiveness and efficiency of schema matching methods for the specific needs of dataset discovery. To this end, we propose \suite, an extensible open-source experiment suite to execute and organize large-scale automated matching experiments on tabular data. \suite{} includes implementations of seminal schema matching methods that we either implemented from scratch (due to absence of open source code) or imported from open repositories. 
The contributions of \suite{} are: $i)$ the definition of four schema matching scenarios as encountered in dataset discovery methods,  $ii)$ a principled dataset fabrication process tailored to the scope of dataset discovery methods and $iii)$ the most comprehensive evaluation of schema matching techniques to date, offering insight on the strengths and weaknesses of existing techniques, that can serve as a guide for employing schema matching in future dataset discovery methods. 
\end{abstract}

\section{Introduction}
Virtually every non-trivial, data science task nowadays begins with data integration. At the core of data integration lies dataset discovery: the process of navigating numerous data sources in order to find relevant datasets as well as the relationships among those datasets. The bulk of work in dataset discovery, focuses on tabular data \cite{cafarella2009data, sarma2012finding, yakout2012infogather,zhang2013infogather+, zhang2017entitables, lehmberg2017stitching, fernandez2018aurum, nargesian2018table, bogatu2020dataset, zhang2020finding, chepurko2020arda} since it constitutes the main form of datasets in the web and enterprises: web tables, spreadsheets, CSV files and database relations. 

Typically, a dataset discovery method receives a dataset as input and finds other datasets in a data repository which are related to it. The ultimate goal of dataset discovery is to augment a dataset with information previously unknown to the user. There are many flavors of dataset discovery: $i)$ searching for tables that can be joined \cite{cafarella2009data, sarma2012finding, lehmberg2017stitching}, $ii)$ augmenting a given table with more data entries or extra attributes \cite{yakout2012infogather, zhang2013infogather+, zhang2017entitables, bogatu2020dataset}, frequently for improving the accuracy of machine learning models \cite{zhang2020finding, chepurko2020arda}, and $iii)$ finding similar tables to a given one using different similarity measures \cite{nargesian2018table, fernandez2018aurum}.

The majority of these methods are based on a common, very critical component: \emph{schema matching}, i.e., capturing relationships between elements of different schemata. In the case of tabular data, dataset discovery methods typically use schema matching techniques to automatically determine whether two columns (or even entire tables) are joinable or unionable. Since dataset discovery methods exploit relatedness information about a given set of datasets, the underlying matching technique of any data discovery method greatly affects its performance.

At the moment of writing, dataset discovery methods typically implement their own matcher, by combining or customizing existing methods. However, the majority of discovery works do not take advantage of the abundance of schema matching methods in the literature \cite{rahm2001survey, do2002comparison}. This happens for good reasons: the vast majority of the techniques are not open-source or available for use, and oftentimes the on-paper description of algorithms can be vague. Worse, most methods require setting a vast number of parameters, making any reproducibility effort a tough or impossible task. Most importantly, even when a few schema matching methods are publicly available, employing them into a dataset discovery pipeline becomes a daunting task: there exists no proper comparison of the state-of-the-art schema matching techniques in the literature -- an open problem which was stated almost two decades ago \cite{rahm2001survey}.

\begin{table*}[t]
\centering
 
 \small
  \begin{tabular}{l||K{1.8cm}|K{2cm}|K{1.8cm}|K{1.5cm}|K{1.8cm}|K{1.8cm}}
  \toprule\bottomrule
    \diagbox{\textbf{Method}}{\textbf{Match Type}}
     & \textbf{Attribute Overlap} \newline \cite{yakout2012infogather,lehmberg2017stitching, bogatu2020dataset} & \textbf{Value Overlap}  \cite{cafarella2009data, yakout2012infogather, lehmberg2017stitching, fernandez2018aurum, nargesian2018table, bogatu2020dataset, chepurko2020arda} &
    \textbf{Semantic Overlap} \newline \cite{sarma2012finding, nargesian2018table} &\textbf{Data Type} \newline \cite{fernandez2018aurum}  &\textbf{Distribution} \newline \cite{fernandez2018aurum,bogatu2020dataset} & \textbf{Embeddings}\newline \cite{nargesian2018table, fernandez2018aurum, bogatu2020dataset} \\
    \toprule\bottomrule
    \textbf{Cupid \cite{madhavan2001generic}} & \checkmark & &\checkmark& \checkmark & &\\
    \hline
    \textbf{Similarity Flooding \cite{melnik2002similarity}} & \checkmark & & &\checkmark & & \\
    \hline
    {\textbf{COMA \cite{do2002coma}}} & \checkmark & \checkmark &
    \checkmark &
    \checkmark & \checkmark&\\
    \hline
    {\textbf{Distribution-based \cite{zhang2011automatic}}} & & \checkmark & & & \checkmark & \\
    \hline
    {\textbf{SemProp \cite{fernandez2018seeping}}} & \checkmark & \checkmark & & & & \checkmark\\
    \hline
    {\textbf{EmbDI \cite{cappuzzo2019local}}} &  &  &  & & & \checkmark \\
    \hline
    {\textbf{Jaccard-Levenshtein}} & & \checkmark & & & & \\
  \toprule\bottomrule
  %
 \end{tabular}
\vspace{1mm}
\caption{\textnormal{Schema matching techniques implemented in  Valentine, and the match types they cover. Match types are marked with the discovery methods requiring them.}}
\label{tab:methods}
\vspace{-6mm}
\end{table*}

In this paper, we present the first work towards evaluating schema matching algorithms on tabular data, for the specific needs of dataset discovery. Traditionally, schema matching algorithms have been evaluated for 1-1 matches: for each column in the source schema, algorithms aim at matching exactly one column in the target schema. This is limiting for dataset discovery use cases where users typically navigate ranked lists of results. We argue that providing ranked lists instead of 1-1 matches, both challenges the traditional matching evaluation metrics (precision and recall), and requires changes to existing algorithms. This work aims to facilitate the development of novel dataset discovery methods by $i)$~automating the schema matching component, $ii)$~by adapting existing algorithms and $iii)$~by proposing novel evaluation metrics with \suite{}: a unified, open-source schema matching experiment suite for dataset discovery.

\noindent The contributions of this paper can be summarized as follows: 
\begin{itemize}
    \item we survey the dataset discovery literature and distill four relatedness scenarios that we strictly define: two joinability and two unionability scenarios;
    \item we \new{extend existing methods} to fabricate dataset pairs for those relatedness scenarios in a principled manner;
    \item we implement and integrate six schema matching algorithms \cite{madhavan2001generic, melnik2002similarity, do2002coma, zhang2011automatic, fernandez2018seeping, cappuzzo2019local} and our own baseline method, and adapt them to the needs of dataset discovery;
    \item we develop a unified and extensible, open-source\footnote{\url{https://github.com/delftdata/valentine}} experimentation suite that can be used as a drop in replacement of the schema matching component in current and future dataset discovery methods;
    \item we present -- to the best of our knowledge -- the most comprehensive effectiveness and efficiency evaluation of schema matching algorithms for tabular data to date, with $\sim$75K experiments (553 dataset pairs $\times$ 135 configurations over multiple schema matching methods).
\end{itemize}

In the rest of the paper we present how schema matching is being used in dataset discovery methods, and propose a new evaluation metric (\autoref{sec:discovery}). We then define a taxonomy with the schema matching scenarios for dataset discovery (\autoref{sec:problemspace}) and how we constructed datasets and ground truth for those cases (Sections~\ref{sec:fabricating-tables} and \ref{sec:datasets}). We then present schema matching methods and the changes required for dataset discovery (\autoref{sec:algorithms}) and finally present experimental results, lessons learned, and open problems (Sections~\ref{sec:findings}, \ref{sec:relwork}, \ref{sec:conclusions}).

\section{From Schema Matching to Dataset Discovery}
\label{sec:discovery}

In this section, we present a concise overview of dataset discovery methods, followed by a discussion on how matching is an integral part of these techniques. Finally, we justify the suitability and necessity of \suite{} as a building block for dataset discovery.

\subsection{Dataset Discovery Methods}
Existing dataset discovery methods on tabular data mainly focus on searching and augmenting/combining information found in related datasets. The early literature in the field has focused on Web Tables and later on dataset repositories. The Octopus system \cite{cafarella2009data} can search and augment Web Tables. It provides the user with three operations: \emph{i)} keyword-search for related datasets, \emph{ii)} specifying semantics of potential new attribute values to a given source, and \emph{iii)} extending data of a given table. InfoGather \cite{yakout2012infogather} and its successor \cite{zhang2013infogather+} introduce methods for augmenting tables either by adding more data entries or by discovering new potential attributes. Similarly, EntiTables~\cite{zhang2017entitables} uses generative probabilistic models in order to augment entity-focused tables, i.e., each row stores information about a specific entity.

In the same spirit, other dataset discovery methods aim specifically at detecting joinable or unionable tables \cite{sarma2012finding, bogatu2020dataset, nargesian2018table} given an input table, often with different end goals, such as improving matching of tabular data to knowledge bases \cite{lehmberg2017stitching}, constructing a knowledge graph to represent relationships between datasets \cite{fernandez2018aurum} or enrich training data and improve accuracy of machine learning methods \cite{chepurko2020arda,zhang2020finding}.

\subsection{The Schema Matching Component}

By studying the literature we observed that the goal of dataset discovery is very similar to the one of schema matching. As a matter of fact, a lot of methods use multiple different matchers in order to identify relationships based on the knowledge sources they have available. For example, if a knowledge base is available and suitable to use then a semantic matcher is used. Furthermore, if a method needs to search for joinable datasets, it might use a matcher that is based on column value overlaps. To help understand the area, we divided those matching needs in six categories as follows (summarized in \autoref{tab:methods}):

\begin{itemize}
    \item \textbf{Attribute Overlap Matcher} (used by \cite{yakout2012infogather,lehmberg2017stitching, bogatu2020dataset}): Specifies that two columns are related when their attribute names have a syntactic overlap above a given threshold.
    \item \textbf{Value Overlap Matcher} (used by \cite{cafarella2009data, yakout2012infogather, lehmberg2017stitching, fernandez2018aurum, nargesian2018table, bogatu2020dataset, chepurko2020arda}): Signals that two columns are related when their corresponding value sets significantly overlap.


    \item \textbf{Semantic Overlap Matcher} (used by \cite{sarma2012finding, nargesian2018table}): In the presence of an external source of knowledge (such as a \emph{knowledge base}), it derives labels describing the semantics of a column or even the domain of its values. Then, a match between two columns is valid when there is a significant overlap between their corresponding labels or, equivalently, they store values of the same domain.
    \item \textbf{Data Type Matcher} (used by \cite{fernandez2018aurum}): Flags (ir)relevant columns based on their data type (integer, string, etc.).
    \item \textbf{Distribution Matcher} (used by \cite{fernandez2018aurum,bogatu2020dataset}): Flags relevant columns based on their value distributions.
    \item \textbf{Embeddings Matcher} (used by \cite{nargesian2018table, fernandez2018aurum, bogatu2020dataset}): Identifies related columns by computing the similarity of their corresponding values based on their embeddings \cite{mikolov2013distributed}. The embeddings are derived from an existing pre-trained model on natural language corpora.
\end{itemize}

Note that it is possible for a given schema matching method to provide more than one type of matchers and, at the same time, a given dataset discovery method might require or use multiple types of matchers. \suite{} encompasses six state-of-the art matching techniques derived from the  schema matching literature plus a baseline approach.  As shown in \autoref{tab:methods}, \suite{}'s' method selection covers all types of matchers used for dataset discovery today.

\para{Valentine as a Discovery Component} \suite{} can contribute to the development of dataset discovery methods in multiple ways. First, it provides a variety of methods for each matcher type, which enables a dataset discovery method to experiment with different techniques based on the data information it can exploit. Moreover, each of \suite{}'s methods includes sophisticated schema matching techniques that cover not one, but several matcher types. In essence, \suite{} consolidates the best of schema matching efforts and make it accessible and usable by dataset discovery methods; Valentine can prevent researchers from having to implement their own, schema matching component or searching through the vast schema matching literature in order to discover techniques well-suited to their needs.

\subsection{Evaluating Matching Techniques for Discovery} 
\label{sec:eval_method}

We use \suite{} to evaluate the performance of multiple schema matching methods by applying them each time on a pair of denormalized tabular datasets with some known schema information - such as table/attribute names and data types - and their associated data values. Moreover, we assume that the intended output consists of matches between columns. An important aspect of the framework is that the output of each method is a list of pairs of matching attributes ranked by the matching confidence as determined by the chosen method. 

\para{1-1 Matches vs. Ranked Matches} Typically, schema matching approaches return a set of 1-1 matches (source to target column matches), however, we argue that rankings are better suited to the needs of dataset discovery: ranking allows users to explore and decide on match candidates more efficiently. Furthermore, it allows us to judge the degree of correctness of a match based on its ranking, thus better reflecting a method's performance. More importantly, it enables dataset discovery methods to utilize these schema matching methods through \suite, since they need to know similarities and rankings among column pairs in order to calculate their corresponding relatedness measures or decide the degree to which two tables can be unioned or joined.

\begin{figure}[t!]
    \centering
    \includegraphics[width=.49\textwidth]{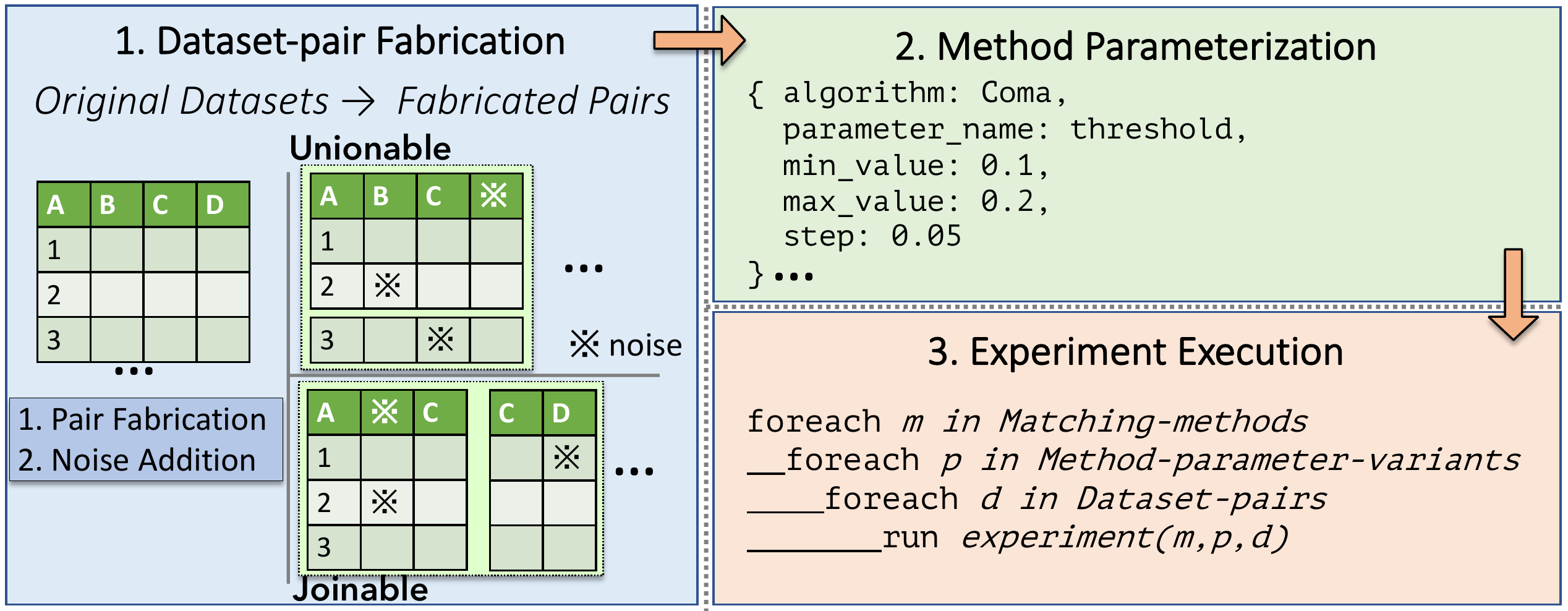}
    \vspace{-5mm}
    \caption{Valentine first fabricates dataset pairs alongside ground truth, then creates multiple parameterized runs of methods and finally exhaustively executes  all combinations of methods, parameters and dataset pairs.}
    \label{fig:pipeline}
    \vspace{-5mm}
\end{figure}

For each pair of relations with potential matches, we know the ground truth, i.e., the matching attribute pairs a schema matching method should capture. This allows us to compute the \new{effectiveness} of each algorithm based on the ranked matches they produced as defined below:

\vspace{-1mm}
\begin{definition}[\textbf{\textit{Recall@\new{ground truth}}}]Measures the number of relevant matches regarding only the top-k match pairs in the result:
    
    \vspace{-2mm}
    \begin{displaymath}
        Recall@\new{ground \ truth} = \frac{\#\  of\ top\text{-}k\ relevant\ matches}{k}
    \end{displaymath}
    where $k = |ground\_truth|$
\end{definition}

\vspace{-2mm}

    
    
    

Recall@\new{ground truth} shows the quality of the ranking a method produces as it computes the top relevant results with respect to the ground truth. Intuitively, it is a measure that reflects how helpful the output list is for a human who wants to assess only a limited list (e.g., a page) of top-$k$ results. In other words, Recall@\new{ground truth} indicates how well a method is able to output all the correct results in the top ranks. Note that since $k = |ground\_truth|$, Recall@\new{ground truth} is essentially equivalent to Precision@\new{ground truth}, hence we only use Recall@\new{ground truth} as an effectiveness metric in this study. 

In our experiments, we exclude traditional effectiveness metrics such as \textit{Precision, Recall and F-measure} since those would apply in the case where matching techniques would return a set of unranked 1-1 matches that satisfy a threshold. \new{While the selected evaluation measure, Recall@ground truth, is not a contribution of our paper, we are the first ones, to the best of our knowledge, to utilize it to evaluate state-of-the-art schema matching methods on their ability to correctly rank matches.}

\section{Dataset Relatedness Scenarios}
\label{sec:problemspace}



\begin{figure}[t!]
    \centering
    \includegraphics[scale=0.4]{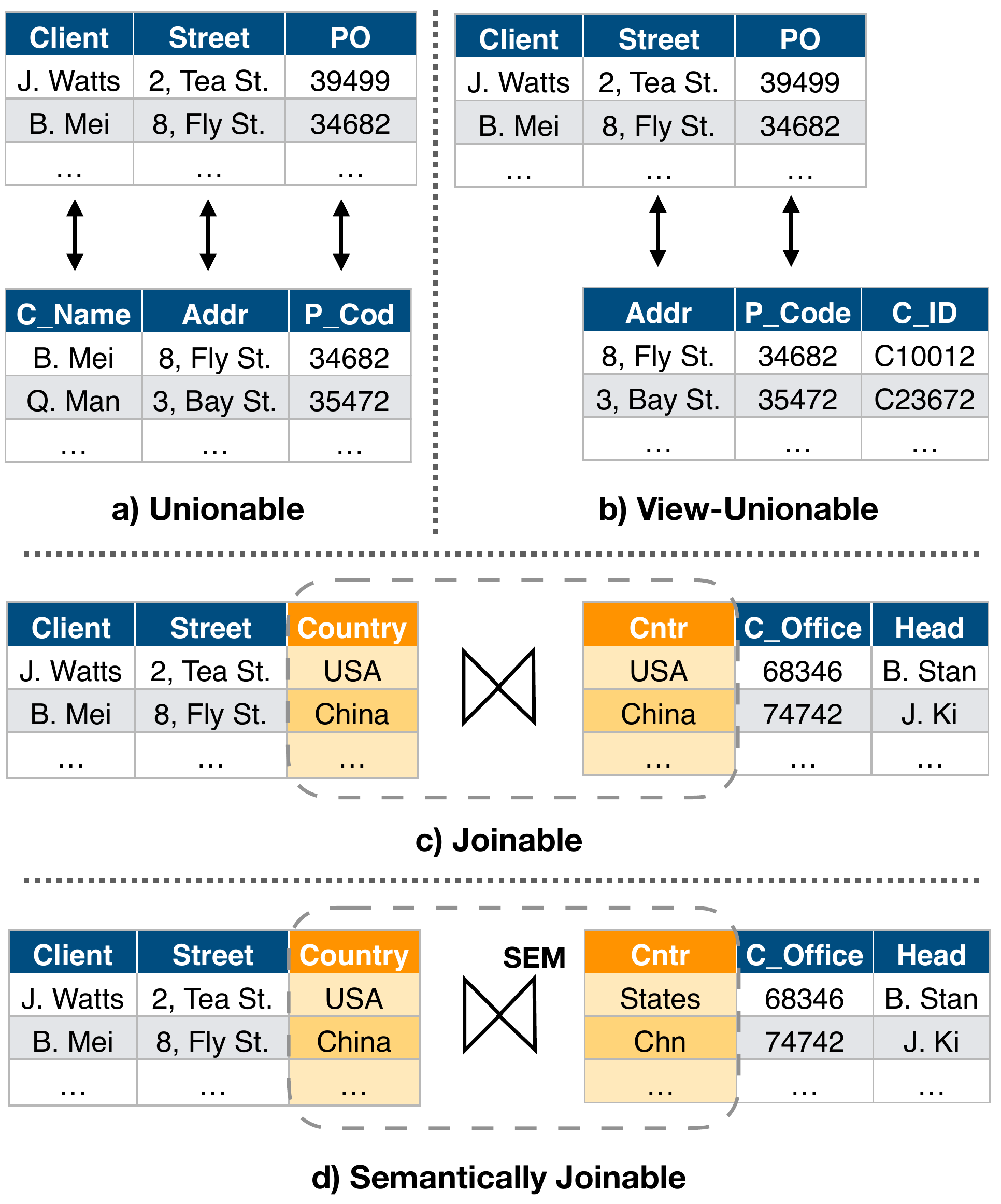}
    \vspace{-3mm}
    \caption{Four cases of dataset relatedness scenarios.}
    \label{fig:prob_space}
    \vspace{-5mm}
\end{figure}

Traditionally, schema matching methods on tabular data are evaluated based on a limited and abstract set of table pairs with a given ground truth of relationships that are valid. However, the scope of a dataset discovery method defines specific relatedness semantics between tables. Therefore, existing schema matching evaluations do not provide any useful insights for dataset discovery techniques.  

In this section, we define and describe the specific relatedness scenarios that \suite{} fabricates in order to meaningfully evaluate existing schema matching methods. Specifically, we develop a relatedness scenario taxonomy with two fundamental categories, \textit{unionable} and \textit{joinable} relations, and further refine each of these categories. This taxonomy covers the scope of \new{recent tabular dataset discovery methods \cite{cafarella2009data, sarma2012finding, yakout2012infogather,zhang2013infogather+, zhang2017entitables, lehmberg2017stitching, fernandez2018aurum, nargesian2018table, bogatu2020dataset, zhang2020finding, chepurko2020arda} }and guides our evaluation in \autoref{sec:findings} as certain approaches can cope with different problem cases better than others.

\color{black}



\subsection{Unionable Relations}
\label{sec:unionable}
In the unionable case, relations store data of the same conceptual entity type using the same attributes. This can be formalized as:


\begin{definition}[\textbf{\textit{Unionable Relations}}]
Two relations $R_1$ with attribute set $\mathcal{A}$ and $R_2$ with attribute set $\mathcal{B}$ are \textbf{unionable} if:
\begin{enumerate}[noitemsep]
    \item They are of the same arity.
    \item There exists a 1-1 mapping $h: \mathcal{A} \rightarrow \mathcal{B}$, denoting semantic equivalence, between their attribute sets ,i.e., $\forall A_i \in \mathcal{A}, \exists B_j \in \mathcal{B}$ so that $h(A_i) = B_j$, and there is no $A_k, k \neq i$ and $B_l, l \neq j$ for which $h(A_k) = B_j$ or $h(A_i) = B_l$.
\end{enumerate}
\end{definition}

Essentially, two relations are unionable if they are \textit{union compatible}, as defined in relational algebra, with the only difference being that corresponding attributes from the two relations may be of different but similar data type (e.g., \textit{string} and \textit{varchar}). This problem can become very challenging when attributes correspond semantically, but their instances mostly differ; yet, a union between the relations should be possible and identifiable. In Figure \ref{fig:prob_space}a we see an example of two unionable relations storing information about clients. Note that even if the names of the corresponding attributes are not the same, they store the same type of information. 

Furthermore, there are a lot of cases where two tables may share a lot of corresponding attributes but also have some extra ones each. This would mean that the two tables are similar but not \textit{unionable}; instead, we call such relations \textit{view-unionable}.

\begin{definition}[\textbf{\textit{View-Unionable Relations}}]
Two relations $R_1$ and $R_2$, with corresponding attribute sets $\mathcal{A}$ and $\mathcal{B}$, are \textbf{view-unionable} if there exist two views $V_1 = \pi_{S_1\subseteq \mathcal{A}}R_1$ and $V_2 = \pi_{S_2\subseteq \mathcal{B} }R_2$, such that $V_1$, $V_2$  are unionable.
\end{definition}

In other words, two view-unionable relations share attributes that correspond to each other semantically, but can also contain attributes that are unique to each; note that in the case where $S_1 \equiv \mathcal{A}$ and $S_2 \equiv \mathcal{B}$ we fall back to the unionable case. This could be a more typical case, since data that is partitioned across different sites, may be differently modelled under the conventions of the respective data owner. More specifically, each such data shard may be enhanced with information (in our case attributes) that are relevant to each owner, thus making it difficult to identify similarity between relations that refer to the same data. An example pair of view-unionable relations is illustrated in Figure \ref{fig:prob_space}b, where we observe that while the two relations share a lot of common attributes, they still differ in the way they refer to clients (one uses names, the other IDs). Thus, they are unionable only with respect to the views defined on their corresponding attributes.

Identification of (view-)unionable relations has been the  goal of several dataset discovery methods \cite{nargesian2018table, fernandez2018aurum} that focus on fetching tables storing similar entities with respect to a given one. Moreover, discovery of unionable relations is vital for techniques that augment information about a given table by finding more data entries to populate it \cite{yakout2012infogather, zhang2013infogather+, bogatu2020dataset}. Thus, Valentine's evaluation on unionable scenarios could be a very important indicator of which existing schema matching methods could effectively enhance such data discovery methods.

\subsection{Joinable Relations}
\label{sec:joinable}

In the joinable case, two relations store complimentary data of the same conceptual entity type. Formally:

\begin{definition}[\textbf{\textit{Joinable Relations}}]
Two relations $R_1$ and $R_2$, with corresponding attribute sets $\mathcal{A}$ and $\mathcal{B}$, are \textbf{joinable} if there exists at least one pair $(A_i, B_j)$, where $A_i \in \mathcal{A}$ and $B_j \in \mathcal{B}$, on which a join can be executed, i.e., $A_i$ and $B_j$ are related through a function $h:\mathcal{A} \rightarrow \mathcal{B}$, which denotes semantic equivalence, and have overlapping instances or $R_1 \bowtie_{A_i = B_j} R_2 \not\equiv R_1 \times R_2$.
\end{definition}


\begin{figure*}
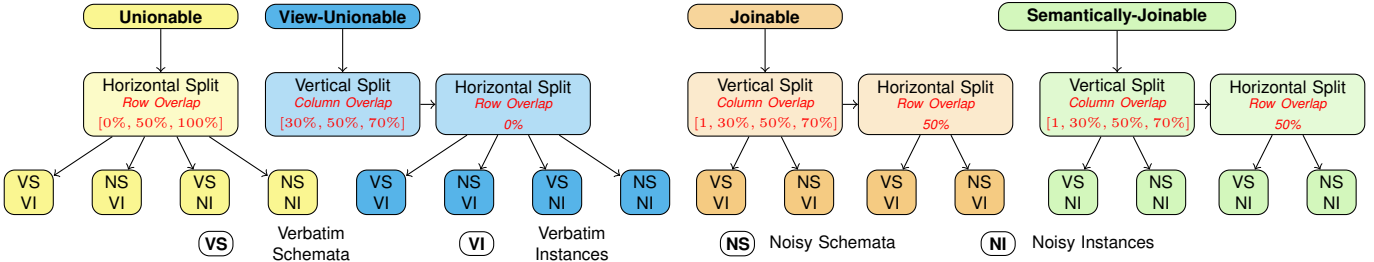

\relscale{.65}
\tikzstyle{nodestyle}=[text width=1.9cm, 
                       align=center, 
                       level distance=.15cm, 
                       font=\sffamily,
                       draw,rounded corners,draw]

\tikzstyle{edgestyle}=[->]

\tikzset{edge from parent/.append style={edgestyle}}

\newcommand{\redtext}[1]{\color{red}\relscale{.8}\textit{#1}}

\hspace*{-1mm}
\begin{tikzpicture}[every node/.style={nodestyle},scale=0.78]
    
    \begin{scope}
    \input{Figures/tikz/unionable-tikz}
    \end{scope}
    
    \begin{scope}[xshift=3.1cm]
    \input{Figures/tikz/v-unionable-itkz}
    \end{scope}
    
    \begin{scope}[xshift=10.3cm]
    \input{Figures/tikz/joinable-tikz}
    \end{scope}
    
    \begin{scope}[xshift=16.3cm]
    \input{Figures/tikz/s-joinable-tikz}
    \end{scope}

\begin{scope}[yshift=2cm]
\matrix [below, draw=none] at (current bounding box.south) {
  \node  [label=right:Verbatim Schemata,text width=3mm] (n1) {\textbf{VS}}; 
  \node  [label=right:Verbatim Instances, text width=3mm, right=30mm of n1] (n2)  {\textbf{VI}};  
  \node [label=right:Noisy Schemata,text width=3mm,  right=30mm of n2] (n3) {\textbf{NS}}; 
  \node [label=right:Noisy Instances,text width=3mm,  right=30mm of n3] (n4) {\textbf{NI}}; \\
};
\end{scope}
\end{tikzpicture}
\vspace{-5mm}
\caption{Fabrication of datasets with respect to each relatedness scenario.}
\label{fig:fabrication-tree}
\vspace{-5mm}
\end{figure*}

Relation joinability can be reduced to finding overlaps between the instance sets of attributes, in the case where data is formatted in the same way for all relations. Figure \ref{fig:prob_space}c shows a classic example of two relations that can join on common values, drawn from the join attributes which are \textit{Country} and \textit{Cntr} respectively. However, capturing joinable relations can become a very hard problem, when they come from diverse data sources. In such cases, it is highly possible that correspondence between instances of two attributes cannot be found due to different format conventions. Therefore, we distinguish this as another joinability problem: one that demands capturing of semantic equivalence.

\begin{definition}[\textbf{\textit{Semantically-Joinable Relations}}]
Two relations $R_1$ and $R_2$, with corresponding attribute sets $\mathcal{A}$ and $\mathcal{B}$, are \textbf{semantically-joinable} if there exists at least one pair $(A_i, B_j)$, where $A_i \in \mathcal{A}$ and $B_j \in \mathcal{B}$ (on which a semantic join can be executed, i.e., $A_i$ and $B_j$ are related through a function $h:\mathcal{A} \rightarrow \mathcal{B}$, which denotes semantic equivalence) share semantically equivalent instances and $R_1 \bowtie^{sem}_{A_i = B_j} R_2 \not\equiv R_1 \times R_2$.
\end{definition}

In essence, semantic-joins are a superset of \textit{fuzzy-joins} \cite{wang2011fast} which have been studied in the literature but only exploit string-based similarities. Figure \ref{fig:prob_space}d showcases the hardness of the problem, where in order to join the two relations, we need a function that captures equivalence between semantically identical values from the \textit{Country} and \textit{Cntr} attributes.

Determining whether relations are (semantically-)joinable is a major necessity for dataset discovery methods that augment a given a table with extra attributes \cite{yakout2012infogather, zhang2013infogather+, bogatu2020dataset}. Moreover, recently, discovery methods search for extra features to augment a given dataset in order to improve accuracy of machine learning models \cite{chepurko2020arda, zhang2020finding}. With our evaluation on joinable scenarios, judging which schema matching method to use in such cases becomes much easier.


\section{Fabricating Dataset Pairs}
\label{sec:fabricating-tables}

Possibly the biggest challenge in evaluating schema matching methods is the lack of openly available datasets with schema matching ground truth. There are \new{three main} ways to create dataset pairs with ground truth: one can $i)$~split existing datasets horizontally to fabricate unionable dataset pairs, and vertically to fabricate joinable dataset pairs \cite{nargesian2018table, zhu2019josie} where the ground truth lies with the original table, $ii)$~curate existing datasets by determining the ground truth manually \cite{zhang2011automatic} or, $iii)$~generate datasets that contain matches by design \cite{alexe2008stbenchmark, arocena2015ibench} (e.g., generate PK-FK relationships). \new{Note that the dataset pairs created by following iii), bear the same characteristics as the datasets that result from splitting them horizontally or vertically using ii), in that they are generated with joinable columns (e.g., by generating intersecting columns). 
In Valentine we opted for i) and ii) as to fabricate dataset pairs from existing datasets and create ground truth. The rest of this section details the fabrication methods.}

\para{Fabricating Dataset Pairs}
\label{sub:synthetic-ground-truth}
We fabricate datasets with synthetic matching challenges by splitting existing tables in a systematic fashion. Here we extend the approach of eTuner~\cite{lee2007etuner} which performs multiple perturbations on the schema and the instances of a table: in short, it splits tables horizontally and vertically, and adds noise in schema information and the value instances. This creates a synthetic matching problem with the original data as ground truth. \new{Moreover, the authors of eTuner showed that using such fabricated datasets to automatically tune schema matching methods leads to better effectiveness on real world datasets than manually tuning them. Therefore, following this approach we are able to represent realistic dataset pair scenarios which lead to robust findings about the effectiveness of the schema matching methods evaluated in our paper.} Below we explain the details of the strategy we followed.

\para{Noise in Data} Apart from keeping the instances of columns \textit{verbatim}
(i.e., after we split a table, we keep the overlapping values the same), we also include \textit{noisy} data in columns as follows: for string columns we insert random typos based on keyboard proximity, while for columns containing only numerical values, we randomly change them according to their value distribution (similar to \cite{lee2007etuner}). 

\para{Noise in Schemata} In the real world, two columns of different tables can have different names, even if they contain the same information. To represent this in our experiments, we include both types of table pairs, i.e., pairs with verbatim column names and pairs in which one of the tables has \textit{noisy} column names. We use a combination of three transformation rules to add ``noise'': $i)$~we prefix column names with their table name (common practice in DB design), $ii)$~we abbreviate column names and $iii)$~ we drop vowels.

We finally split tables horizontally to create unionable pairs, vertically to create joinable pairs, and in both ways (joinable and unionable), following ~\cite{lee2007etuner, nargesian2018table}. \autoref{fig:fabrication-tree} shows the dataset fabrication process for four relatedness scenarios (\autoref{sec:problemspace}).

\para{Unionable} To create datasets for the \textit{unionable} case we need two tables to contain the same columns. Thus, we horizontally partition the table with varying percentages of row overlap, which is necessary for instance-based matching methods. As mentioned above, such a table pair might contain verbatim schemata or noisy ones, as well as verbatim or noisy instances. We use all possible instances-schemata combinations, while the ground truth for each case consists of \textit{all} corresponding columns of the two horizontally-split tables that match.

\newcommand{\texta}{Verbatim Instances (VI)}
\newcommand{\textb}{Noisy Inst\-ances (NI)}
\newcommand{\textcn}{Verbatim Schemata (VS)}
\newcommand{\textdn}{Noisy Schemata (NS)}


\para{View-unionable} For the \textit{view-unionable} case, we need two tables with a common subset of columns, but no row overlap. This represents a typical matching problem in practical applications, i.e., finding more instances of a given type scattered across tables with slightly varying schema representation. The lack of row overlap provides an extra challenge for naive instance-based algorithms. We create \textit{view-unionable} cases by splitting the original table both horizontally and vertically with zero row overlap and varying column overlap. Again, we consider every feasible instances-schemata combination.

\para{Joinable} \textit{Joinable} tables should have at least one (joining) column in common and, in contrast to view-unionable, they should have a large row overlap. This represents the common challenge of finding additional information/features about known data instances in other tables. To create this case, we split a table vertically keeping a varying amount of overlapping columns (e.g., 1 column, or 30\% of columns or 50\%, etc.). Another way to create \textit{joinable} tables is to split the table both vertically and horizontally but with a row overlap of different percentage (in our case 50\%). We create variants with noise/no-noise in each schema, but since we refer to the ``classical'' join operation we include only verbatim instances. 

\para{Semantically-joinable} The \textit{semantically-joinable} case is similar to the \textit{joinable} case, but we perturb the overlapped instances by inserting noise. Thus, because of noise, an equality join on the common columns will not yield the original table anymore. As before, we create variants with noise/no-noise in the schema, but include only \textit{noisy} instances (non-noisy instances are the ``vanilla'' \textit{joinable} case). 

\section{Datasets}
\label{sec:datasets}

We have selected a set of datasets to evaluate the schema matching methods (see Section~\ref{sec:algorithms}) included in Valentine.
The datasets bear distinct characteristics such that they challenge all methods.
We group the datasets in two broad categories.
The first category presented in Section~\ref{sub:fabricated-ground-truth} contains dataset sources that provided us with a total of 540 fabricated dataset pairs by applying Valentine's fabricator module on them as we described in Section~\ref{sec:fabricating-tables}.
In this case the ground truth are the original tables.
The second category presented in Section~\ref{sub:curated-ground-truth} features real-world datasets with an inherent schema matching challenge that we curated in order to manually create the ground truth for them.


\subsection{Dataset Sources of Fabricated Dataset Pairs}
\label{sub:fabricated-ground-truth}

\para{TPC-DI \cite{tpcdi} - 180 pairs} TPC-DI focuses on Data Integration. We used the \textit{Prospect} table from TPC-DI 1.1.0 with a scale factor of three. The fabricated TPC-DI datasets vary from 11 to 22 columns and 7492 to 14983 rows.

\para{Open Data~\cite{nargesian2018table} - 180 pairs} This dataset consists of tables from Canada, USA and UK Open Data, provided to us by the authors of \cite{nargesian2018table} for their dataset discovery techniques. We used the second table from the \texttt{base.sqlite} collection of the benchmark. \rev{The fabricated Open Data datasets vary from 26 to 51 columns and 11628 to 23255 rows.}

\para{ChEMBL\footnote{\url{https://www.ebi.ac.uk/chembl/}} - 180 pairs} ChEMBL is an open chemical database closely related to the \textit{EFO\footnote{\url{https://www.ebi.ac.uk/efo/}}} ontology. Thus, it is one of the few datasets that come with an ontology. We used the \textit{Assays} table from ChEMBL 22. \rev{The fabricated ChEMBL datasets vary from 12 to 23 columns and 7500 to 15000 rows.}

\subsection{Dataset Sources of Human-curated Dataset Pairs}
\label{sub:curated-ground-truth} 

\para{WikiData\footnote{\url{https://www.wikidata.org}} - 4 pairs} WikiData is a knowledge base supporting Wikimedia projects and is a great source of real world data. We create two tables as a matching challenge covering the same entity type queried from WikiData, but represented with slightly varying schemata and instance encodings. We focus on singers who are USA citizens. The schemata for these tables are identical at first: both cover twenty columns containing mostly strings (e.g. artist name, parents name, song genre). To resemble a real-life scenario as accurately as possible, we vary the column names of the second table (e.g. partner $\rightarrow$ spouse). Additionally, we change the values for all cells of six selected columns by replacing the original value with alternative versions (e.g., Elvis Presley $\rightarrow$ Elvis Aaron Presley). Finally, we manually created variants for all matching classes of the matching scenarios as in the previous subsection, \rev{with relations varying from 13 to 20 columns and 5423 to 10846 rows}.

\para{Magellan Data \cite{magellandata} - 7 pairs} The Magellan Data Repository \cite{magellandata} contains dataset pairs collected from real-world data and curated mainly for Entity Matching techniques. We pick 7 of these datasets pairs which have been previously used for Schema Matching evaluation in \cite{cappuzzo2019local}. With respect to our relatedness scenarios, the datasets represent unionable pairs of tables with value overlaps and use the same naming conventions between corresponding columns. Magellan datasets vary from 3 to 7 columns and 864 to 131099 rows.

\para{ING Data (proprietary) - 2 pairs} Our industry partner ING Bank Netherlands provided us with access to two production datasets, comprising a pair of matching tables each. The first pair of tables (ING\#1) contains information about SCRUM sprints with dates, team ids, owner-team, tasks, EPIC names, dates, etc. The bank owns multiple custom SCRUM systems that they would like to integrate and query for team-performance analysis. \rev{The corresponding tables consist of 33 columns - 935 rows and 16 columns - 972 rows respectively.}

The second dataset (ING\#2) contains tables that describe the software applications that a team is responsible for, alongside information like the owner-team, the hardware it operates on, the manager name, department, the relationships between applications (e.g., app1 is used by app2), etc. The dataset contains two tables: a wide one (\rev{with 59 columns - 1000 rows}) with low-level general-domain information, and another (\rev{with 25 columns - 1000 rows}) containing higher-level business-oriented information. These tables are denormalized, and even contain nested/composite values. Finding matches in this dataset is very challenging also for human domain experts, and semi-automated matching for cases like this would be very appreciated by practitioners. Thus, this dataset is a very good test case for schema matching methods.

We gathered the ground truth for both datasets with the help of an expert DB admin who performed the schema matching manually. Unfortunately, we cannot make this dataset public due to privacy constraints.


\section{Matching Methods}
\label{sec:algorithms}

Schema matching approaches are classified based on the kind of information they make use of. In specific, schema-based matching methods \cite{madhavan2001generic, melnik2002similarity, do2002coma} exploit only schema-level knowledge in order to capture potential relationships, such as attribute names, data types and contextual information. On the other hand, instance-based matching approaches rely on data instances, such as those that compare value distributions of attributes \cite{zhang2011automatic} or compute various syntactic similarity measures \cite{engmann2007instance}. Finally, there exist hybrid methods that combine both schema and value information \cite{fernandez2018seeping, cappuzzo2019local}. In this section we give a brief overview of each method contained in Valentine, and explain our parameter configuration process.

\subsection{Methods Description}

In what follows we briefly describe the schema matching methods that we either integrated or implemented in \suite. Furthermore, we explicitly report any modifications we made while attempting to reproduce the original algorithms. 

\para{Cupid \cite{madhavan2001generic}} Cupid is a schema-based approach. 
Schemata are translated into tree structures representing the hierarchy of different elements (relations, attributes etc.). The overall similarity of two elements is the weighted similarity of i) \textit{Linguistic Matching} and ii) \textit{Structural Matching}. The first calculates the name similarity for each pair of elements from the two schemata belonging to the same \textit{category}. Structural matching utilizes the tree transformations of the schemata to compute similarity between elements based on their context. The overall similarity of two elements is the weighted sum of the linguistic and structural similarities. Cupid is not openly-available, thus in our implementation we used \textit{WordNet}\footnote{\url{https://wordnet.princeton.edu/}} as thesaurus, while we rely on the name similarity formula to compute data compatibility scores.



\para{Similarity Flooding \cite{melnik2002similarity}} Similarity Flooding is a schema-based matching approach that relies on graphs, and outputs correspondence between any kind of elements (relations, attributes, data types) of two given schemata. Specifically, the schemata are transformed to directed graphs, which have as nodes every element and as edges the relationships that these elements have with each other (e.g. a relation has an attribute, which is of a certain type). The graphs are then merged into a \textit{propagation graph}, where pairs of nodes having similar connections collapse into \emph{map pairs}. The intuition of the algorithm is that each such map pair propagates its similarity to its neighbors, causing an update in their similarity score in an iterative manner, until convergence. In our study we have implemented from scratch the original method (since there exists only an outdated Java version of it from 2003), with the only difference that we use a string similarity of our own choice, i.e. \textit{Levenshtein distance} \cite{levenshtein1966binary}, since there are no details on the actual function that the authors used.

\para{COMA \cite{do2002coma}} COMA combines multiple schema-based matchers. Schemata are represented as rooted directed acyclic graphs, where the associated elements are graph nodes connected by edges of different types (e.g. containment). The match result is a set of element pairs and their corresponding similarity score. COMA also supports human feedback by allowing users to indicate the correctness of the resulting matches, which is taken into consideration in next iterations, allegedly improving general accuracy. \cite{engmann2007instance} extended COMA to also incorporate two instance-based matchers. In our experiments we use the COMA 3.0 Community Edition, where we use the default schema-based and instance-based strategies.




\para{Distribution-based Matching \cite{zhang2011automatic}} Distribution-based Matching is an instance-based method. Relationships between different columns are captured by comparing the distribution of their respective data values. The method computes and refines clusters  of relational attributes, using the \textit{Earth Mover's Distance} (EMD) 
between pairs of columns, which is a measure of distribution similarity of the corresponding instance sets. In the end, a number of disjoint clusters is given as output, wherein relational attributes are considered to be related. We implemented the original method (which was not openly-available) without any modifications, except for using another software for solving the integer programming problem in the last step of the algorithm, which decides the final clusters (we used \textit{PuLP}\footnote{\url{https://pythonhosted.org/PuLP/}} instead of \textit{IBM CPLEX}).



\para{SemProp \cite{fernandez2018seeping}} SemProp tries to capture relationships between schema elements beyond syntactic similarity by making use of pre-trained \textit{word embeddings} \cite{mikolov2013distributed}. SemProp first builds a \textit{semantic matcher} that given 
a domain-specific ontology links attribute and table names to ontology classes using their embedding representation; then it relates disparate attributes and tables by transitively following these links. Pairs of elements that fail to be related by the semantic matcher are forwarded to a syntactic one. In our experimental evaluation, we make use of the open-sourced code for the \textit{Aurum}
\cite{fernandez2018aurum} dataset discovery system, which includes the SemProp matcher and a domain-specific ontology to make the method run on the ChEMBL datasets.


\para{EmbDI \cite{cappuzzo2019local}} EmbDI is a framework facilitating data integration tasks on relational data, by building \emph{relational embeddings}. The authors propose a method for embedding values and attribute names of relations, by training them based on the input without using pre-trained embeddings. However, the method uses external knowledge, such as synonym dictionaries or pre-trained embeddings, in order to deal with more challenging cases. EmbDI is eligible for schema matching tasks, where it finds relationships between the columns of two datasets by comparing their corresponding embeddings. We integrated EmbDI in \suite{} by importing the code\footnote{\url{https://gitlab.eurecom.fr/cappuzzo/embdi}} accompanying the original paper.

\para{Jaccard-Levenshtein Matcher} As a simple baseline, we implemented a naive instance-based matcher computing all pairwise column similarities by using Jaccard similarity. We treat two values as being identical if their Levenshtein distance is below a given threshold. The method outputs a ranked list of column pairs, along with their respective similarity score. 


\subsection{Method Parameterization}
\label{sec:experiments}

\begin{table}[t!]
\centering
 \smaller
\resizebox{\columnwidth}{!}{
\begin{tabular}{l|c||c||c}
\toprule\bottomrule
\textbf{Method} & \textbf{Parameter} & \textbf{Values} & \textbf{Step} \\ \toprule\bottomrule
\multirow{3}{*}{\textbf{Cupid \cite{madhavan2001generic}}} & \texttt{leaf\_w\_struct} & \texttt{{[}0, 0.6{]}} & \texttt{0.2} \\ \cline{2-4} 
 & \texttt{w\_struct} & \texttt{{[}0, 0.6{]}} &\texttt{0.2} \\ \cline{2-4} 
 & \texttt{th\_accept} & \texttt{{[}0.3, 0.8{]}} & \texttt{0.1} \\ \hline
\multirow{2}{*}{\textbf{Sim. Fl. \cite{melnik2002similarity}}} & \texttt{prop.coeff.} & \texttt{inverse\_average} & \texttt{-} \\\cline{2-4} 
 & \texttt{fix-point comp.} & \texttt{C} & \texttt{-} \\ \hline
\multirow{2}{*}{\textbf{COMA \cite{do2002coma}}} & \texttt{strategy} & \texttt{{[}schema, inst.{]}} & - \\ \cline{2-4} 
 & \texttt{threshold} & \texttt{0} & \texttt{-} \\ \hline
\multirow{2}{*}{\textbf{Dist.\#1 \cite{zhang2011automatic}}} & \texttt{phase 1   $\theta$} & \texttt{{[}0.1, 0.2{]}} & \texttt{0.05} \\ \cline{2-4} 
 & \texttt{phase 2 $\theta$} & \texttt{{[}0.1, 0.2{]}} & \texttt{0.05} \\ \hline
\multirow{2}{*}{\textbf{Dist.\#2 \cite{zhang2011automatic}}} & \texttt{phase 1 $\theta$} & \texttt{{[}0.3, 0.5{]}} & \texttt{0.1} \\ \cline{2-4} 
 & \texttt{phase 2 $\theta$} & \texttt{{[}0.3, 0.5{]}} & \texttt{0.1} \\ \hline
\multirow{3}{*}{\textbf{SemProp \cite{fernandez2018seeping}}} & \texttt{minh.threshold} & \texttt{{[}0.2, 0.3{]}} & \texttt{0.1} \\ \cline{2-4} 
 & \texttt{sem.threshold} & \texttt{{[}0.4, 0.6{]}} & \texttt{0.1} \\ \cline{2-4} 
 & \texttt{coh.sem.threshold} & \texttt{{[}0.2, 0.4{]}} & \texttt{0.2} \\ \hline
 \multirow{4}{*}{\textbf{EmbDI \cite{cappuzzo2019local}}} & \texttt{train. algorithm} & \texttt{word2vec} & \texttt{-} \\ \cline{2-4} 
 & \texttt{sentence\_length} & \texttt{60} & \texttt{-} \\ \cline{2-4} 
 & \texttt{window\_size} & \texttt{3} & \texttt{-} \\ \cline{2-4}
 & \texttt{n\_dimensions} & \texttt{300} & \texttt{-} \\ \hline
\textbf{Jacc. Lev.} & \texttt{threshold} & \texttt{{[}0.4, 0.8{]}} & \texttt{0.1} \\ \toprule\bottomrule
\end{tabular}
}
\vspace{1mm}
\caption{\textnormal{Parameterization of implemented matching methods. For each parameter combination we run a separate experiment, as shown in \autoref{fig:pipeline}.}}
 \label{tab:param_config}
\vspace{-6mm}
\end{table}

\new{In order to ensure that we conduct a fair evaluation of each schema matching method, we configure each method using the parameters provided by the authors of Similarity Flooding \cite{melnik2002similarity}, COMA~\cite{do2002coma}, and EmbDI~\cite{cappuzzo2019local} in the corresponding papers. Unfortunately, this is not possible for Cupid~\cite{madhavan2001generic}, Distribution-based \cite{zhang2011automatic}, SemProp \cite{fernandez2018seeping} and our Jaccard-Levenshtein baseline method because default parameter values are not available. Instead, we perform a grid search for these methods and datasets at hand as shown in \autoref{tab:param_config} in order to discover the parameter values resulting to the best performance.} The parameters that are not included are set to their default values as described in the respective papers. We performed two different runs for the distribution-based method \cite{zhang2011automatic}. The first based on the recommended threshold values of the original paper, and the second to help the method find more matches in column pairs with low overlap. Additionally, we split the single global threshold that was proposed in two, one for each phase. For COMA, we allow the output to include any found element pair, regardless of their similarity (i.e. we set the \textit{accept similarity threshold} parameter to be 0). Finally, in Cupid we ran experiments with the weight of the structural similarity \emph{w\_struct $\leq$ 0.6}, since tabular data do not have the complex structure of XML schemata for which the method was designed. 

\subsection{Sensitivity to Parameter Changes under Grid Search}

To evaluate the sensitivity of each of the four methods' effectiveness we performed grid search with respect to parameter variations. More specifically, we vary a single parameter value ceteris paribus and apply each method to all 180 dataset pairs of the CheMBL database. We do this for all the different values that we vary per parameter, as illustrated in \autoref{tab:param_config}. Note that CheMBL is the only dataset source on which all four methods can be applied.
We measure the effectiveness sensitivity by means of standard deviation from the mean effectiveness score (i.e. recall at ground truth) for each dataset pair. Finally, we compute the overall minimum, median, and maximum standard deviation with respect to each parameter, and present them in \autoref{tab:param_sens}.
In the interest of space, we include only the parameters taking at least 3 different values.

We make two notable observations. The minimum and median standard deviation for all methods is close to zero, which means that the change of parameter value had practically zero effect to the methods' effectiveness. This could be explained by a high instance overlap and value or attribute similarity, which allows a method to capture relevance regardless the configuration.
On the contrary, the maximum standard deviation is considerable (close to 0.5) for most of the parameters. 
It shows that methods can be very sensitive with respect to the thresholds and weights they use in order to assess whether a similarity score can be accepted as an indication of matching columns. This is especially observed when dataset pairs share few values or contain noise and thresholds are low, implying as a rule of thumb that a stricter threshold can drastically lead to better results.


\begin{table}[t!]
\centering
 \smaller
\resizebox{\columnwidth}{!}{
\begin{tabular}{l|c||c|}
\toprule\bottomrule
\textbf{Method} & \textbf{Varying Parameter} & \textbf{St. Deviation (Min, Median, Max)} \\ \toprule\bottomrule
\multirow{3}{*}{\textbf{Cupid \cite{madhavan2001generic}}} & \texttt{leaf\_w\_struct} & \texttt{{[}0, 0.04, 0.43{]}} \\ \cline{2-3} 
 & \texttt{w\_struct} & \texttt{{[}0, 0.04, 0.5{]}}\\ \cline{2-3} 
 & \texttt{th\_accept} & \texttt{{[}0, 0.05, 0.5{]}} \\ \hline
\multirow{2}{*}{\textbf{Dist-based \cite{zhang2011automatic}}} & \texttt{phase 1 $\theta$} & \texttt{{[}0, 0, 0.47{]}} \\ \cline{2-3} 
 & \texttt{phase 2 $\theta$} & \texttt{{[}0, 0, 0.14{]}} \\ \hline
\textbf{SemProp \cite{fernandez2018seeping}} & \texttt{sem.threshold} & \texttt{{[}0, 0, 0.08{]}} \\ \hline
\textbf{Jacc. Lev.} & \texttt{threshold} & \texttt{{[}0, 0.04, 0.49{]}} \\ \toprule\bottomrule
\end{tabular}
}
\vspace{1mm}
\caption{\textnormal{\new{Impact of parameters of schema matching methods expressed through min, median and max standard deviation values across all ChEMBL datasets.}}}
 \label{tab:param_sens}
\vspace{-6mm}
\end{table}

\color{black}

\begin{figure*}[t!]
    \centering
    \includegraphics[scale=0.28]{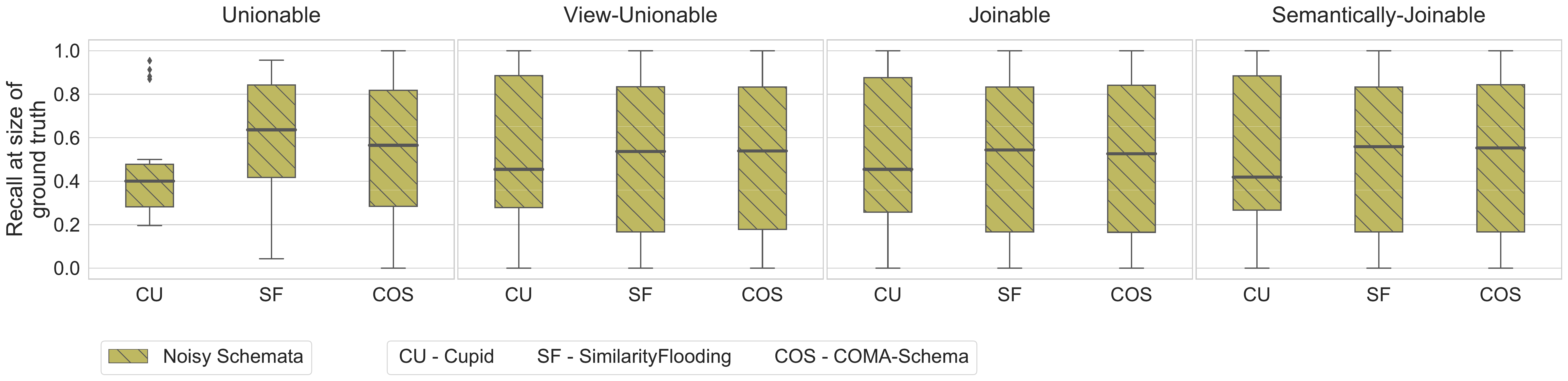}
    \vspace{-3mm}
    \caption{Effectiveness results of Valentine's schema-based matching methods for each dataset relatedness scenario}
    \label{fig:schema}
    \vspace{-0.25cm}
\end{figure*}

\section{Findings}
\label{sec:findings}
We assess the performance of schema matching methods through an exhaustive set of experiments, as shown in \autoref{fig:pipeline}. In the following, we summarize \rev{the effectiveness} of all matching methods measured by \new{\rgt{}} (\autoref{sec:eval_method}) over all conducted experiments showing minimum, median and maximum \rgt{} values. \new{In specific, each box shows the range of \rgt{} values that a method exhibits when tested against several fabricated dataset pairs, which adhere to the relatedness scenarios discussed in \autoref{sec:problemspace}}. Furthermore, we assess the efficiency of the approaches by presenting the average execution time of each matching method over all dataset pairs. An extensive collection of all detailed experimental results per dataset source can be found in our code repository. 

\begin{figure*}[t!]
    \centering
    \includegraphics[scale=0.28]{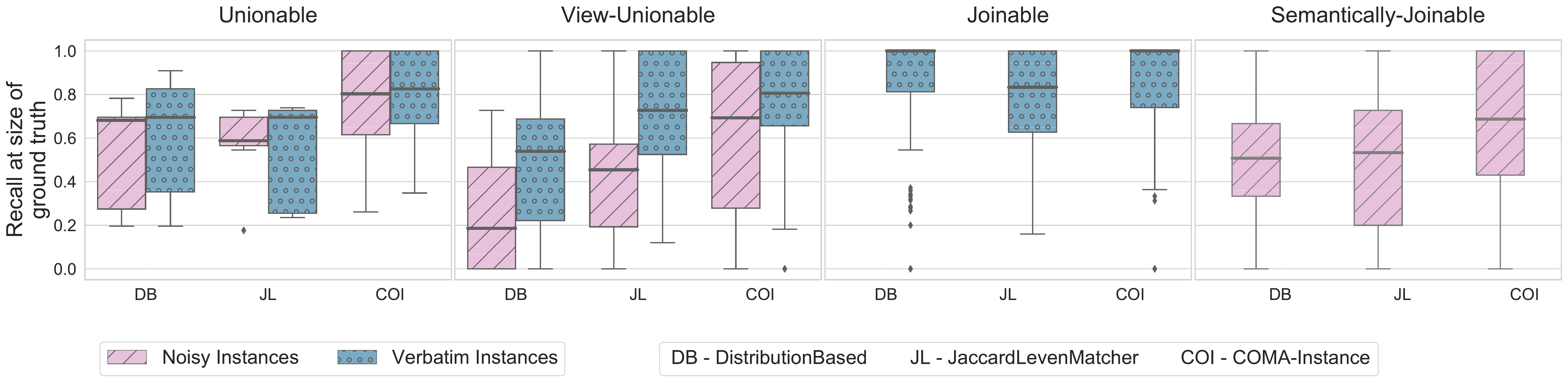}
    \vspace{-3mm}
    \caption{Effectiveness results of instance-based matching methods for each dataset relatedness scenario.}
    \label{fig:instance}
    \vspace{-4mm}
\end{figure*}


\vspace{-2mm}
\subsection{Fabricated Dataset Pairs (TPC-DI, Open Data, ChEMBL)}

\label{sec:synthetic_findings}
\subsubsection{Schema-based Methods}
In \autoref{fig:schema} we focus on methods that leverage only schema-level information, such as attribute names and data types: Cupid \cite{madhavan2001generic}, Similarity Flooding \cite{melnik2002similarity} and the schema-based flavor of COMA \cite{do2002coma}. 
First, we see that when matching columns are represented by different attribute names, because of noise that we have introduced in the schemata, there is no schema-based method that can provide satisfying and consistent results in any scenario. Specifically, 
we see that Similarity Flooding and COMA outperform Cupid, yet their effectiveness is varying with median \rgt{} close to 0.6. 
To summarize, in the absence of good attribute names, the rest of the schema information graph (e.g., types, transitive relationships) or contextual information such as the neighborhood of columns per dataset do not actually give any useful insights for any schema-based method.


\subsubsection{Instance-based Methods} 

\autoref{fig:instance} shows effectiveness results for Valentine's instance-based methods, which only exploit the corresponding value sets of each dataset's columns: Distribution-based matching \cite{zhang2011automatic}, the instance-based flavor of COMA \cite{engmann2007instance} and our Jaccard-Levenshtein baseline. The first interesting observation is that the view-unionable relatedness scenario is considerably harder than the unionable one. The main reason for this is that there are extra vertical splits on the tables, and there is no row-overlap to help instance-based matchers. 
In addition, all instance-based methods show worse results for semantically-joinable datasets compared to the joinable ones. This is a consequence of the dissimilarity between the instance sets of corresponding attributes. The high dispersion in effectiveness and significantly lower median \rgt{} values that even state-of-the-art methods provide regardless the sophisticated similarity measures they use point to a valuable take-away message: capturing semantic similarity between relations with respect to their corresponding instances is a hard problem. In fact, all methods output results with high skew in effectiveness (except for the joinable scenarios), which proves that we are comfortably far from ``out of the box'' instance-based matchers.

\begin{figure*}[t!]
    \centering
    \includegraphics[scale=0.28]{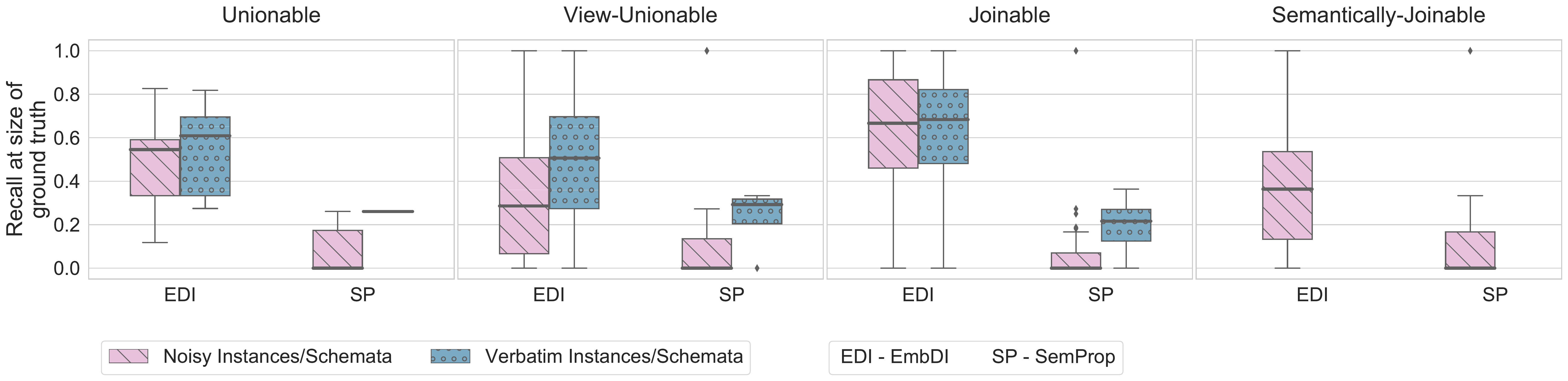}
    \vspace{-3mm}
    \caption{Effectiveness results of hybrid matching methods for each dataset relatedness scenario.}
    \label{fig:hybrid}
    \vspace{-4mm}
\end{figure*}
\begin{figure*}[t!]
    \centering
    \includegraphics[scale=0.32]{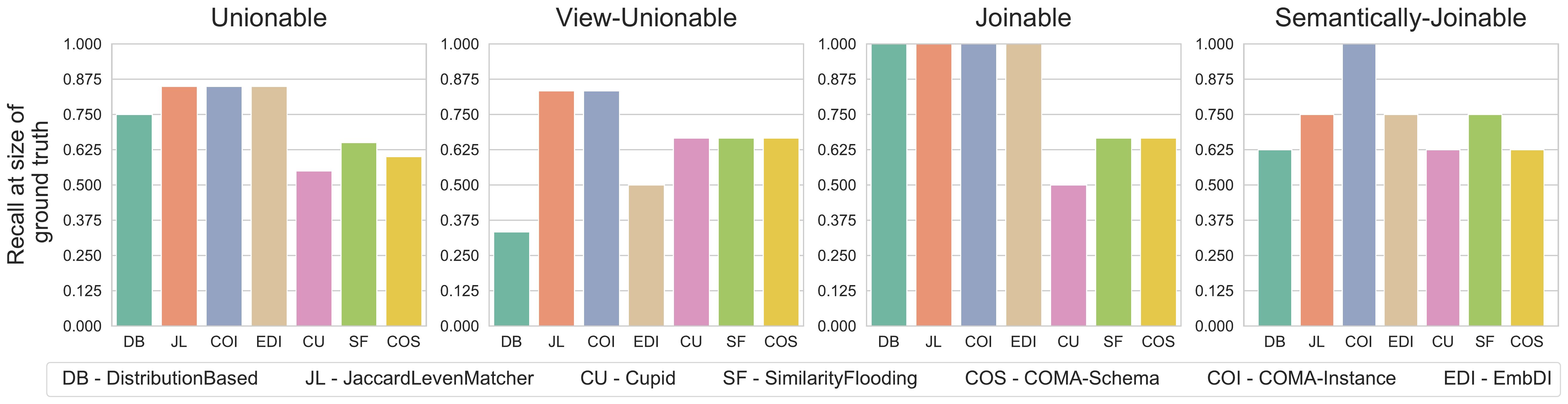}
    \vspace{-3mm}
    \caption{\rev{Effectiveness} results on WikiData.}
    \label{fig:wiki}
    \vspace{-0.45cm}
\end{figure*}

\subsubsection{Evaluation of Hybrid Methods} In \autoref{fig:hybrid} we summarize results for Valentine's hybrid matching methods, which utilize both schema and instance-level information: EmbDI \cite{cappuzzo2019local} and SemProp \cite{fernandez2018seeping}. 
To begin with, SemProp's effectiveness is unexpectedly low over all relatedness scenarios, worse than any other matching method we tested with Valentine. Therefore, we observe that the pre-trained word embeddings that SemProp leverages in order to capture relatedness are not reliable, since they cannot help when the data domain is too specific (as in the case of ChEMBL data).
On the other hand, EmbDI is more effective than SemProp, but it provides with inconsistent and low \rgt{} values across all dataset pairs. This is particularly unexpected for dataset pairs that are semantically-joinable, since we would anticipate that the local embeddings of EmbDI will be able to capture semantics of data instances better than any other matcher. However, it performs the worst among all schema- and instance-based methods, due to the randomness in training data generation and the dependence on overlapping instance values; in the case where the overlapping values are few or missing, and external knowledge is absent, the method struggles to accurately capture context and semantics of data elements.

\subsubsection{Expected Results}

In \autoref{fig:schema} we opted for showing results only for noisy schemata, since we verified that with verbatim schemata all schema-based methods are able to place all correct matches at top. Furthermore, in \autoref{fig:instance} we see that instance-based methods perform better in the absence of noisy instances, especially in the case of joinable dataset pairs; columns that can be joined share the same instances. For the same reason, in \autoref{fig:hybrid} we see that EmbDI provides acceptable results in the case of joinable scenarios.

\subsection{Human-Curated Dataset Pairs (WikiData, Magellan, ING)}


\subsubsection{WikiData}

In Figure \ref{fig:wiki} we see the \new{effectiveness} results for the dataset pairs coming from WikiData (see \autoref{sec:datasets}). First, all four instance-based methods exhibit better \rgt{} than the schema-based ones, in \textit{unionable} relations. This is reasonable, since they can leverage the overlaps of the corresponding attributes' instance sets, while the schema-based ones heavily rely on attribute names which, in some cases, are very different. For \textit{view-unionable} relations, we observe similar behavior, but with a major difference: distribution-based matching gives results of poor quality due to discrepancy in value distributions. This is owed to the fabrication of matching columns with varying distribution similarity (using horizontal splits and by adding noise). 

The instance-based methods are able to find all relevant matches and place them in the top ranks (\rgt{}=1), when the relations are \textit{joinable} due to high data value overlaps. In contrast, schema-based methods are unable to find some of the correct matches, since they can only exploit attribute names and types. The COMA instance-based approach is the clear winner in the case of \textit{semantically-joinable} relations, being able to provide every correct match in the first places of the ranked list even in the existence of noise. 
Moreover, the difference in the names of corresponding columns makes it even more difficult for schema-based approaches to perform as expected. This confirms again that in all considered scenarios, the instance-based techniques are superior to the schema-based ones. 


\subsubsection{Magellan Data}

\autoref{tab:ing_results1} summarizes the effectiveness of Valentine's matching methods over all dataset pairs drawn from the Magellan data repository as discussed in \autoref{sec:datasets}. With the exception of COMA, all other methods that use instance information are not able to have the same effectiveness as Valentine's schema-based approaches, which leverage the fact that matching columns have the same attribute names. This mainly happens due to minor discrepancies between value sets of matching columns, which as we saw in \autoref{sec:synthetic_findings} complicates the effectiveness of instance-based or hybrid matching methods. Furthermore, Magellan datasets may contain multi-valued attributes (such as lists of actors for movie datasets) that add extra complexity. 
As a final remark, we see that the results we get from Magellan Data are not as informative as the ones we got from our fabricated dataset pairs. Conversely, they provide no or misleading information on the advantages or disadvantages of the different schema matching method categories, while they do not cover all our relatedness scenarios which are highly important and relevant for any dataset discovery approach. 
\begin{table}[b]
\vspace{-4mm}
\small
\begin{tabular}{l|c||c||c}
\toprule\bottomrule
\textbf{Methods} & \textbf{Magellan} & \textbf{ING\#1} & \textbf{ING\#2} \\ \toprule\bottomrule
\textbf{Cupid \cite{madhavan2001generic}} & \textbf{1} & 0.714 & 0.5 \\ \hline
\textbf{Similarity Flooding \cite{melnik2002similarity}} & \textbf{1} & 0.357 & 0.439 \\ \hline
\textbf{COMA Schema-based \cite{do2002coma}} & \textbf{1} & 0.786 & 0.121 \\ \hline
\textbf{COMA Instance-based \cite{massmann2011evolution}} & \textbf{1} & 0.786 & 0.136 \\ \hline
\textbf{Distribution-based \cite{zhang2011automatic}} & 0.54 & \textbf{0.857} & \textbf{0.879} \\ \hline
\textbf{Jaccard Levenshtein} & 0.787 & 0.786 & 0.621 \\  \hline 
\textbf{EmbDI \cite{cappuzzo2019local}} & 0.818 & 0.714 & 0.227\\ \toprule\bottomrule
\end{tabular}
 \vspace{1mm}
 \caption{\textnormal{Recall at size of ground truth for the Magellan and ING Data.}}
 \label{tab:ing_results1}
\vspace{-3mm}
\end{table}
\subsubsection{ING Data}
In \autoref{tab:ing_results1} we summarise the performance of the seven methods upon the two provided backlog datasets from ING, as described in \autoref{sec:datasets}.

\para{ING\#1} For the first dataset we expected the schema-based algorithms to perform better than the instance-based ones. This is because the corresponding/matching columns between the two tables have either identical or very similar names. At the same time, the corresponding columns contain hashes, descriptions and similar words that are used in multiple contexts (i.e., can create false positives). Contrary to our expectations, almost all of the methods managed to find around  70\% of the expected matches, with the exception of Similarity Flooding which placed a lot of false positives in the top ranks. 
The Distribution-based method performed the best, one of the reasons being that these tables contained a lot of almost-identical values in the matching columns, leading to very similar distributions that created matches. Interestingly, the Jaccard-Levenstein method could find most of the matches, but with some false-positives ranked high. The reason is that Jaccard-Levenshtein does not compare distributions but actual set similarity measures. 

\para{ING\#2} Our expectation for this dataset was that schema-based algorithms would not perform well, as the column names of the second table, contained suffixes that could complicate schema-based-matching. On the other hand, we expected that instance-based methods would work well: the instances in dataset ING\#2 were even more similar than the ones of ING\#1. Moreover, the ground truth contained multiple matches for each column of the small table to lots of columns of the 60-column table. The Distribution-based method performed far better than any other algorithm for similar reasons to the ones we outlined above. On the other hand, COMA, although we configured it to match each source-column with more than one target-column, it did not find a lot of those target-column matches. We believe that to be a bug of the current version of COMA (v3.0). Finally, we see that EmbDI's local embeddings could not accurately capture relationships between matching columns, since the randomness that inhibits in the method's training set construction does not facilitate capturing relevance.

\begin{table}[t]
\centering
\begin{tabular}{l|c}
\toprule\bottomrule
\textbf{Methods} & \textbf{Average Runtime} \\ 
\toprule\bottomrule
\textbf{Cupid \cite{madhavan2001generic}} & 9.64 \\ \hline
\textbf{Similarity Flooding \cite{melnik2002similarity}} & 7.09 \\ \hline
\textbf{COMA Schema-based \cite{do2002coma}} & 1.67 \\ \hline
\textbf{COMA Instance-based \cite{massmann2011evolution}} & 318.07 \\ \hline
\textbf{Distribution-based \cite{zhang2011automatic}} & 71.16 \\ \hline
\textbf{SemProp \cite{fernandez2018seeping}} & 735.25 \\ \hline
\textbf{EmbDI \cite{cappuzzo2019local}} & 4817.87 \\ \hline
\textbf{Jaccard Levenshtein} & 522.94 \\ \toprule\bottomrule
\end{tabular}
\vspace{2mm}
\caption{\textnormal{Average runtime per experiment (i.e., table pair) in seconds.}}
\label{tab:efficiency}
\vspace{-7mm}
\end{table}

\subsection{Efficiency Results}
We executed all experiments as batch jobs in two 80-core Linux virtual machines, with 320 GB of RAM each; experiments on the ING datasets ran on our partner's in-house machines for privacy reasons, hence they are excluded. In \autoref{tab:efficiency} we show the average runtime per method over all dataset pairs. First of all, we see that schema-based methods are by far the most efficient since they avoid looking into instance values; Cupid and Similarity Flooding are considerably slower than COMA due to the fact that they build and process structures that attempt to exploit context (trees and graphs respectively). 

On the other hand, methods that utilize instance-level information are several orders of magnitude slower, with EmbDI exhibiting the worst runtime overall. Specifically, we observed that EmbDI's bottleneck is the random walk generation part which does not scale efficiently when the number of available instances grow; in addition, the training of embeddings can be very time consuming. Furthermore, we observe that the Distribution-based and COMA are the most efficient instance-based methods. Nonetheless, we noticed that both of them can exhibit very long execution times, mainly due to heavy processing they apply on data values, where COMA invokes procedures on sets of values and the Distribution-based method applies a two-stage clustering.

\section{Related Benchmarks and Frameworks}
\label{sec:relwork}

The \textit{Ontology Alignment Evaluation Initiative} (OAEI)  \cite{euzenat2011ontology, euzenat2013ontology} encompasses a multitude of benchmarks related to matching with respect to ontologies. However, it does not cover methods capturing relevance among tabular data. STBenchmark \cite{alexe2008stbenchmark} focuses on the evaluation of \textit{mapping systems}, which specify logical assertions to exchange data between a source and a target.
Its evaluation is guided on a defined set of schema mapping scenarios that are orthogonal to  
\suite{}, which is to the best of our knowledge the first study devoted to the dataset discovery literature.

iBench~\cite{arocena2015ibench} presents a metadata generator for data integration tasks targeting large and complex schema mappings. By opposite,  eTuner~\cite{lee2007etuner} provided a method for automatically tuning schema matching systems; Valentine builds upon the ideas behind dataset and ground truth generation from eTuner (\autoref{sec:fabricating-tables}). The only closest attempt to ours is presented in XBenchMatch~\cite{duchateau2007xbenchmatch}, which  evaluated schema matching tools by focusing only on XML data, and not on tabular data.  Moreover, it included a few datasets and methods in the analysis and it was not designed for schema matching for the primary needs of dataset discovery.
\vspace{-2mm}

\section{Conclusion \& Lessons learned}
\label{sec:conclusions}
Our work was motivated by the lack of a comprehensive experimental framework to compare the performance and effectiveness of existing schema matching techniques as core operations for dataset discovery. To the best of our knowledge, this paper contributes the first comprehensive and large-scale experiment suite, encompassing over 500 dataset pairs, state of the art schema matching tools and meaningful dataset discovery scenarios. To stimulate further research, \suite~is entirely open-source (including data, ground truths, scenarios, outputs) and easily reproducible. 
Our analysis led to a number of lessons learned as discussed below. 

\para{One size does not fit all} Our evaluation over both Valentine's fabricated dataset pairs and those stemming from real-world data show that there is not a single schema matching method that consistently performs better than others. Instead, we see that COMA \cite{do2002coma}  exhibits higher effectiveness over most of our fabricated dataset pairs, yet the Distribution-based method \cite{zhang2011automatic} is the most well-suited for our real-world ING datasets. Consequently, we believe that following COMA's approach of \textit{composing} state-of-the-art matching methods (e.g., by adding the recent embeddings-based approaches), should be the preferred way in dataset discovery or other integration pipelines.

\para{Embeddings for matching} Our experimental results showed that SemProp's pre-trained embeddings provide with low effectiveness when used in isolation. On the other hand, EmbDI's local embeddings can improve effectiveness, yet most of the times they do not perform as well as other state-of-the-art schema or instance-based methods. Therefore, while we acknowledge that embeddings-based techniques can improve effectiveness by incorporating them into existing matching methods, we believe that further research is needed in order to make them effective.

\para{Complex parameterization} Most methods require complex parameterization in order to perform well. 
For the most part, parameters are dependent on the input data that needs to be matched, which makes it very hard for practitioners to use those methods. We believe that our community should focus on  ``self-driving'' matching methods that do not require parameterization \cite{lee2007etuner}. Machine learning might be a solution to some of the parameterization problems \cite{dong2018data}, but then would require at least some availability of ground truth to steer the learning process. The experimental results presented in this work represent idealized near-optimal conditions as we determined most parameters by performing a grid-search (which exploited our ground truth). In the wild, we expect to see lower performance for most algorithms as parameters are then likely not optimized.

\para{Simple baselines perform well} Our simple baseline Jaccard-Levenshtein matcher (ca. 70 lines of Python code) works surprisingly well, especially considering its simplicity. We argue that similar baselines to ours, along with the rest of the methods discussed in this paper, can foster future comparative analysis for schema matching and dataset discovery processes.

\para{Humans-in-the-loop} ``Self-driving'' matching methods should be able to work alongside humans giving feedback on the matching process, not in the form of parameters or thresholds, but in the form of positive/negative examples, etc. In the same spirit, the design of schema matching methods should focus on presenting matches as ranked candidates; we strongly believe that the schema matching problem should be approached as a \textit{search problem}, rather than an \textit{optimization problem} (e.g., find the best set of 1-1 matches of columns). Schema matching of the future should focus more on preparing results that will be shown to humans, and should utilize feedback from humans \cite{li2017human}. 


\para{Schema Matching is resource-expensive} Instance-based methods are still expensive as they have to calculate similarity metrics between large sets where it can be very expensive to find matches. Future research should focus on approximate methods to allow for better scaling \cite{zhu2016lsh, zhu2019josie, fernandez2019lazo}.


\balance

\bibliographystyle{IEEEtran}
\bibliography{references} 

\balance

\end{document}